\documentclass{article}

\usepackage{PRIMEarxiv}

\usepackage[utf8]{inputenc} 
\usepackage[T1]{fontenc}    
\usepackage{hyperref}       
\usepackage{url}            
\usepackage{booktabs}       
\usepackage{amsfonts}       
\usepackage{nicefrac}       
\usepackage{microtype}      
\usepackage{lipsum}
\usepackage{fancyhdr}       
\usepackage{graphicx}       
\graphicspath{{media/}}     
\usepackage{float}
\usepackage{tabularx}
\title{Deep Reputation Scoring in DeFi: zScore-Based Wallet Ranking from Liquidity and Trading Signals}

\author{
Dhanashekar Kandaswamy\thanks{All authors contributed equally to this work.} \\
The Ohio State University, USA \\
\texttt{kandaswamy.6@buckeyemail.osu.edu}
\and
Ashutosh Sahoo \\
Zeru Finance, Panama \\
\texttt{ashutosh@zeru.finance}
\and
Akshay SP \\
Zeru Finance, Panama \\
\texttt{akshay@zeru.finance} \\
\and
Gurukiran S \\
Zeru Finance, Panama \\
\texttt{gurukiran@zeru.finance}
\and
Parag Paul \\
Rig AI, USA \\
\texttt{parag@rigai.co}
\and
Girish G N \\
Zeru Finance, Panama \\
\texttt{girish@zeru.finance}
}

\begin{document}
\maketitle

\begin{abstract}
As decentralized finance (DeFi) protocols continue to reshape digital asset ecosystems, the need to distinguish between different types of user behavior—liquidity providers vs. active traders—has become crucial for risk modeling, incentive alignment, and on-chain reputation. In this paper, we present a comprehensive behavioral scoring framework tailored to Uniswap, the leading decentralized exchange (DEX). Our system assigns two complementary scores: a \textbf{Liquidity Provision Score} that quantifies the strategic depth and stability of a user's liquidity contributions, and a \textbf{Swap Behavior Score} that captures trading intent, volatility exposure, and transactional discipline.

We first construct blueprint-based rule systems that decompose each score into interpretable components, including volume, frequency, holding time, and withdrawal patterns. To handle edge cases and learn cross-feature interactions, we introduce a noise-injected refinement phase using a deep residual neural network composed of dense blocks with skip connections, inspired by U-Net architecture. Importantly, we incorporate \textbf{pool-level context}—such as total value locked (TVL), fee tiers, and relative pool size—enabling the model to differentiate identical user behavior across pools of varying scale and risk.

Our method enables nuanced, context-aware scoring of DeFi users, providing a scalable backbone for credit risk assessment and incentive design in on-chain systems. We validate our framework on Uniswap v3 transaction data and demonstrate its potential to power data-driven segmentation and protocol-aligned reputation mechanisms.

\textit{Although we refer to our behavioral metric as \textbf{zScore}, it is independently developed and methodologically distinct from the cross-protocol reputation system introduced by Udupi et al.~[16]. Our approach focuses on role-specific behavioral modeling within Uniswap, combining blueprint-driven score generation with supervised learning.}

\end{abstract}

\section{Introduction}

The emergence of decentralized finance (DeFi) has fundamentally reshaped the financial landscape, replacing traditional intermediaries and centralized trust mechanisms with transparent, algorithmically governed, and permissionless protocols \cite{malamud2017decentralized, lo2021uniswap}. One of the most impactful developments in this space is decentralized exchanges (DEXs), platforms enabling peer-to-peer asset swaps, liquidity provisioning, and trading through smart contracts. Unlike centralized exchanges, which mandate custodial control, extensive identity verification, and regulatory oversight \cite{schueffel2019evaluating, fantazzini2021crypto}, DEXs operate pseudonymously, identifying users solely by their on-chain activity. This structural distinction grants significant freedom and autonomy but simultaneously introduces critical challenges—particularly in user segmentation, incentive alignment, and behavioral assessment.

Uniswap, the leading DEX, exemplifies these transformative shifts through its automated market maker (AMM) mechanism, enabling permissionless liquidity provision and token swaps without intermediaries \cite{lo2021uniswap, aigner2021uniswap}. Despite its revolutionary approach to decentralized trading and capital efficiency, Uniswap and similar protocols face substantial limitations due to the absence of identity-based frameworks \cite{lin2024riskprop, nguyen2024reputation}. Traditional credit scoring systems, reliant on verified identities, banking histories, or centralized credit bureaus, become fundamentally inadequate within these pseudonymous, blockchain-based environments \cite{packin2024decentralized, shimpi2024credit}. As a result, evaluating a user's trustworthiness, strategic alignment, or long-term credibility becomes challenging, relying solely on transactional histories—deposits, withdrawals, token swaps, and holding behaviors.

This limitation underscores a significant gap in current DeFi risk assessment and user modeling approaches. Most existing protocols overly rely on simplistic heuristics such as total transaction volume or collateral ratios, often neglecting the nuanced behavioral traits of wallet activity \cite{mussoi2025risk, heimbach2022risks}. Additionally, existing scoring systems frequently exhibit rigidity due to fixed thresholds and lack adaptability to dynamic user behaviors or market conditions \cite{carter2021defi, doerr2021defi}. Moreover, opaque and non-interpretable machine learning methods exacerbate the challenges of fairness, bias, and transparency, creating black-box models that are difficult for users and protocol developers alike to trust \cite{bucker2022transparency, provenzano2020machine}.

Recognizing these shortcomings, our work seeks to answer an essential question: Can a wallet's behavior, purely derived from on-chain interactions, serve as a robust basis for evaluating its reputation and reliability within DeFi ecosystems? We argue affirmatively, proposing a novel dual-role behavioral scoring framework that evaluates liquidity providers (LPs) and active traders (swappers) separately, leveraging transaction-level interaction patterns from Uniswap v3.

Our methodology first constructs an interpretable \textit{behavioral blueprint}, systematically decomposing user actions into quantifiable dimensions such as deposit frequency, holding duration, withdrawal strategies for LPs, and transaction intensity, asset diversification, and rebalancing frequency for traders. This initial heuristic-driven scoring, while transparent and controllable, inherently faces limitations: fixed decision boundaries, oversensitivity to minor behavior changes, and lack of flexibility in dynamic market scenarios.

To overcome these constraints, we introduce a hybrid scoring architecture that integrates a noise-augmented, deep residual neural network with skip-connected dense blocks, inspired by U-Net-style architectural principles. This model preserves interpretability by starting from rule-based, heuristic-generated scores and enhances adaptability through supervised learning that captures complex, nonlinear feature interactions. Crucially, our scoring approach incorporates contextual pool-level metrics, such as total value locked (TVL), fee tier, and relative pool scale, ensuring that identical actions are appropriately weighted based on the environmental context \cite{heimbach2022risks, aigner2021uniswap}.

Our primary contributions can be summarized as follows:
\begin{itemize}
\item We propose a dual-track, behavior-based scoring system tailored specifically to liquidity providers and traders, addressing critical gaps in user segmentation and credit risk assessment on Uniswap v3.
\item We develop a two-phase scoring pipeline combining a transparent, heuristic-based behavioral blueprint with a noise-injected deep residual network, balancing interpretability and expressiveness.
\item We explicitly integrate pool-level contextual awareness into our scoring mechanism, enabling nuanced differentiation of identical behaviors across varying market environments.
\item We validate our comprehensive framework using real-world Uniswap v3 transactional data, demonstrating its utility for credit risk modeling, reward structuring, lending protocols, and strategic user segmentation.
\end{itemize}

Ultimately, by transforming complex and rich on-chain behavioral data into structured and interpretable signals, our framework bridges the existing gaps between anonymous DeFi interactions and meaningful, trust-based protocol alignment, maintaining the foundational values of decentralization, openness, and autonomy.

\section{Related Work}

\subsection{Scoring and Reputation in Crypto Systems}

Initial research into crypto scoring primarily concentrated on centralized exchanges. Schueffel and Groeneweg \cite{schueffel2019evaluating} developed a multi-criteria model to evaluate centralized exchanges using indicators like trust, fees, and user support. Fantazzini and Calabrese \cite{fantazzini2021crypto} leveraged machine learning techniques and survival analysis to forecast centralized exchange closures, emphasizing regulatory and operational risk factors. Although foundational, these approaches primarily focus on platform-level assessments and do not address individual user behavior in decentralized finance contexts.

\subsection{Wallet-Level Risk and Reputation Models}

With decentralized finance shifting the emphasis towards wallet-level activity, several recent models assess individual wallet risk and reputation. Lin et al. \cite{lin2024riskprop} proposed RiskProp, a de-anonymization-based network propagation technique to evaluate wallet risk, particularly for identifying malicious actors. Udupi et al. \cite{udupi2025zscore} introduced zScore, a universal reputation system leveraging semi-supervised neural networks to assign cross-protocol scores based on user activity. While valuable in scope, such systems often rely on inferred or partially labeled data, which can present challenges when fine-grained distinctions are required—e.g., separating wallets scoring 850 vs. 900 in the absence of explicit labels.

In contrast, our framework adopts a blueprint-guided scoring paradigm. We first define an interpretable, domain-informed behavioral rubric, and then inject controlled noise to generate smooth, structured targets for supervised training. This allows us to model nuanced behavioral differences while avoiding hard classification decisions on unlabeled users. While both systems aim to quantify wallet-level trust and behavior, our approach is independently developed and designed specifically for the decentralized exchange (DEX) environment, with an emphasis on role-specific behavior (e.g., liquidity provision vs. trading).

Nguyen et al. \cite{nguyen2024reputation} developed a lending-focused reputation model employing the PageRank algorithm to quantify borrower trustworthiness. While valuable, their method lacks detailed granularity at the behavioral level necessary for evaluating liquidity providers or active traders. Mussoi et al. \cite{mussoi2025risk} compared risk forecasting methods in AMM protocols through macro-level time series analysis, without translating these insights into actionable, wallet-level behavioral scoring.

Additionally, Jain et al. \cite{jain2024wire} introduced "Wire," a Web3 integrated reputation engine that combines machine learning with on-chain behavioral metrics to score user credibility. Packin and Lev-Aretz \cite{packin2024decentralized} critically analyzed decentralized credit scoring systems, highlighting issues of opacity and fairness when incorporating on-chain behaviors into reputation metrics.

\subsection{Market Microstructure, LP Risks, and DEX Behavior}

Several studies have extensively explored the structural and behavioral dynamics of decentralized exchanges and liquidity providers. Malamud and Rostek \cite{malamud2017decentralized} established theoretical underpinnings for decentralized exchange mechanisms using general equilibrium frameworks. Lo and Medda \cite{lo2021uniswap} analyzed Uniswap’s automated market maker (AMM) structure, providing foundational insights into permissionless liquidity and trade execution dynamics. Aspris et al. \cite{aspris2021decentralized} characterized decentralized exchanges as volatile and minimally regulated marketplaces, highlighting inherent risks and uncertainties. Hägele \cite{hagele2024centralized} systematically compared centralized and decentralized exchanges, emphasizing their differing approaches to transparency, incentives, and governance structures.

Focused studies on liquidity provider risks underscore the complexity of behavioral dynamics within AMM-based platforms. Heimbach et al. \cite{heimbach2022risks} empirically analyzed risks and returns for Uniswap v3 liquidity providers, demonstrating significant variation based on provider strategy and market volatility. Similarly, Aigner and Dhaliwal \cite{aigner2021uniswap} quantified impermanent loss, underscoring the necessity of risk-adjusted metrics for liquidity provision strategies. Carter and Jeng \cite{carter2021defi} and Doerr et al. \cite{doerr2021defi} highlighted systemic risks within DeFi protocols, critiquing existing approaches for insufficiently addressing dynamic user behaviors and market responsiveness.

\subsection{Machine Learning Approaches to Credit and Behavioral Scoring}

The application of machine learning in credit scoring has seen significant advancements, primarily due to its capacity to capture complex, non-linear relationships within user data. Dastile et al. \cite{dastile2020statistical} provided a comprehensive review emphasizing transparency, feature importance, and interpretability in credit scoring models. Dumitrescu et al. \cite{dumitrescu2022machine} explored enhancements to logistic regression via decision-tree-derived non-linear effects, emphasizing improved prediction accuracy and interpretability. Bücker et al. \cite{bucker2022transparency} specifically addressed the importance of transparency and auditability in machine learning-based credit scoring, underscoring the critical balance between performance and model interpretability.

\subsection{Positioning Our Contribution}

Our research explicitly addresses gaps identified in the existing literature by introducing a dedicated behavioral scoring pipeline tailored specifically for DEX users on Uniswap v3. Unlike existing frameworks, our approach distinctly evaluates liquidity provisioning and trading behaviors through a dual-role system. Importantly, our scoring process begins with a structured, interpretable behavioral blueprint that captures high-level traits such as transaction frequency, holding duration, and asset diversity, rather than relying on inferred or weak labels.

We then apply a hybrid scoring strategy that enhances adaptability without sacrificing transparency: a blueprint-guided scoring layer is followed by a supervised deep residual network trained on noise-augmented labels, enabling the model to generalize across behavioral patterns. This architecture avoids the pitfalls of black-box inference while capturing cross-feature interactions in a principled manner.

Finally, our scoring mechanism incorporates contextual pool-level variables—such as total value locked (TVL), fee tiers, and relative pool scale—to differentiate otherwise similar user actions based on market environment. To our knowledge, this is the first interpretable and context-aware behavioral scoring system purpose-built for decentralized exchange ecosystems.

\section{Methodology}

Our methodology begins with the collection of raw on-chain data from Uniswap v3, a decentralized exchange protocol deployed on the Ethereum blockchain. These transaction logs are publicly accessible, immutable, and verifiable, making them a reliable source for analyzing user behavior in decentralized finance. The data is retrieved using subgraph queries and blockchain APIs that expose smart contract event logs associated with user actions. We focus on three primary transaction types: liquidity deposits, liquidity withdrawals, and token swaps. Deposit events capture the addition of assets to a specific liquidity pool, while withdrawal events reflect the removal of liquidity and, optionally, the collection of accrued fees. Swap events record asset exchanges within a pool and include information about the input and output tokens, route complexity, and associated pool parameters such as fee tier. Each transaction is annotated with metadata including wallet address, transaction hash, timestamp, token symbols and amounts, and pool-specific identifiers. This dataset forms the foundation of our behavioral analysis pipeline and enables a structured understanding of user activity across multiple pools and timeframes.

\subsection{Feature Blueprint}

To ensure transparency and interpretability in our scoring system, we construct a domain-informed feature blueprint that outlines which behavioral traits influence a user's score—and in what proportion. Rather than relying on hard-labeled datasets or unsupervised clustering, our blueprint provides a soft, role-aware scoring schema grounded in DeFi mechanics. This allows us to encode prior knowledge directly into the structure of the scoring function, enabling consistent interpretation and modular refinement.

The blueprint is organized into two parallel structures, corresponding to liquidity providers (LPs) and swap participants, each tailored to capture role-specific behavioral patterns.

For liquidity providers, the blueprint captures dimensions such as deposit volume and frequency, withdrawal activity, and average holding time. These features reflect not only capital commitment but also the temporal discipline with which liquidity is managed. We additionally incorporate wallet age and LP volatility—the latter quantifying the consistency of provisioning patterns over time. While no rigid cutoffs are enforced, each trait is assigned a relative weight, allowing users to achieve high scores only through consistently balanced behavior. Pool-level context is embedded as well: fee tier impacts cost exposure, and TVL shapes the influence of behavior across high- and low-liquidity pools.

For traders, the blueprint evaluates engagement through swap volume and frequency, while capturing risk posture and strategic behavior via token diversity, volatility exposure, and average time between swaps. Additional behavioral signals include routing complexity (as a proxy for sophistication) and integrity indicators that penalize micro-swaps, wash trading, or low-effort patterns. Similar to LPs, contextual factors like fee tier and TVL are incorporated at the aggregation stage.

Overall, the blueprint acts as a structured scaffold for generating interpretable, role-specific behavioral scores. These blueprint-derived signals serve not as hard labels but as soft, behavior-informed training targets. This approach allows us to balance interpretability with adaptability, and avoids the brittleness of hand-tuned thresholds or error-prone manual classification.

\subsection{Proposed Methodology}

Our behavioral scoring system is designed as a hybrid two-phase architecture that blends interpretability with adaptability. The objective is to produce wallet-level scores that reflect not only individual behavioral traits but also the behavioral environment of the pools each user interacts with. Rather than relying on fully opaque learning systems, our approach is grounded in a domain-informed scoring blueprint that guides the model’s understanding of strategic behavior while enforcing fairness and role-specific balance.

\paragraph{Blueprint-Guided Score Initialization.}
The first phase of the pipeline generates soft, interpretable training targets based on the behavioral blueprint. Each wallet is assigned an initial behavior-based score derived from contributions across fixed sub-categories—such as deposit consistency, holding time, or swap diversity. Each sub-category is capped by a predefined maximum, ensuring that no single trait dominates the overall score. This weighting strategy enforces balance: high scores must result from consistently disciplined behavior across multiple dimensions. While inherently interpretable, this rule-based approach may suffer from rigidity, especially near decision boundaries or for edge-case behaviors.

\paragraph{Noise-Aware Label Generation.}
To mitigate brittleness and promote smoother generalization, we inject controlled Gaussian noise into the blueprint-generated scores before training. These noise-augmented targets preserve the original blueprint's ordinal structure and sub-category contributions while introducing soft boundaries. This helps the model learn a flexible, differentiable mapping between user behavior and scores, without relying on brittle rule cutoffs. The result is a training signal that is rooted in domain understanding, yet capable of modeling behavioral nuance.

\paragraph{Neural Model Architecture.}
\begin{figure}[ht]
    \centering
    \includegraphics[width=\linewidth]{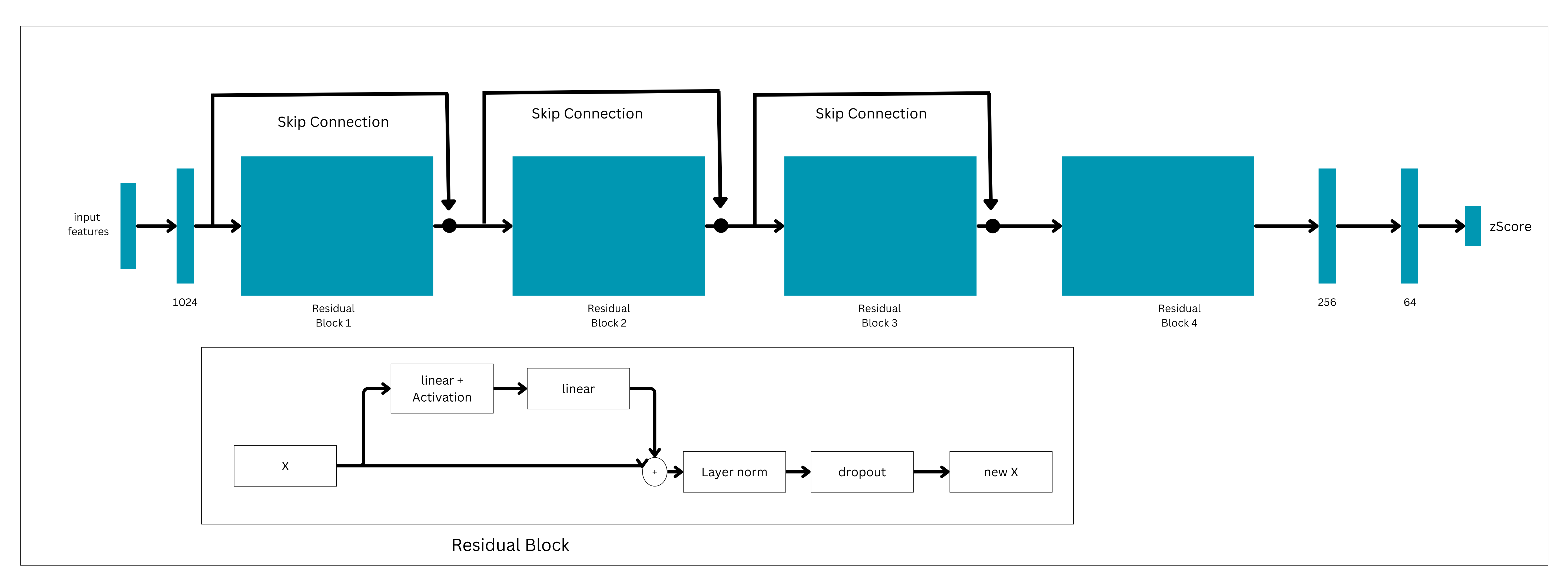}
    \caption{
        \textbf{Model Architecture.} 
        The top portion illustrates the full DeepMLP\_ResNet architecture, which processes input features through an initial projection layer followed by four residual blocks with skip connections. Dimensionality is progressively reduced from 1024 to 256 across the blocks. The final regression head maps the processed representation to a scalar zScore. The bottom inset zooms into a single residual block, showing its internal structure: two linear layers (with activation), residual addition, layer normalization, and dropout. Projection shortcuts are used when input and output dimensions differ.
    }
    \label{fig:model_architecture}
\end{figure}

In the second phase, we train a supervised regression model to approximate and refine the noisy blueprint scores. We employ a deep fully connected network with residual connections, termed \textbf{DeepMLP\_ResNet}, which supports both model depth and stable optimization. The architecture begins with an input projection layer, followed by four residual blocks that enable deep feature composition while maintaining gradient flow through skip connections.

Each residual block contains two linear layers with SiLU activation, followed by a residual addition, layer normalization, and dropout. When input and output dimensions differ, the skip path includes a projection shortcut to ensure compatibility before addition. The model progressively reduces dimensionality across residual layers, using the following transitions: $1024 \rightarrow 512 \rightarrow 256$.

After the final residual block, the output is passed through a two-layer regression head:
\begin{itemize}
    \item \texttt{Linear(256 → 64)} → LayerNorm → SiLU
    \item \texttt{Linear(64 → 1)} to produce the final scalar score
\end{itemize}

The final output is a continuous behavioral score trained to align with the noisy blueprint-derived targets. This architecture allows the model to generalize across behavioral heterogeneity while preserving the interpretability and structure of the original blueprint.

\paragraph{Training Setup and Objective.}
We train the model using the AdamW optimizer with a learning rate of $5 \times 10^{-4}$ and weight decay of $10^{-4}$, minimizing mean squared error (MSE) loss. A ReduceLROnPlateau scheduler is used to adjust the learning rate based on training loss. The model is trained for 500 epochs with early stopping based on validation criteria. To prevent information leakage, we enforce wallet-level train/validation splits before feature aggregation, ensuring that no wallet appears in both sets. All features are normalized using z-score scaling.

\section{Results and Findings}
\subsection{LP Score}
\begin{figure}[H]
    \centering
    \begin{minipage}[b]{0.48\textwidth}
        \centering
        \includegraphics[width=\textwidth]{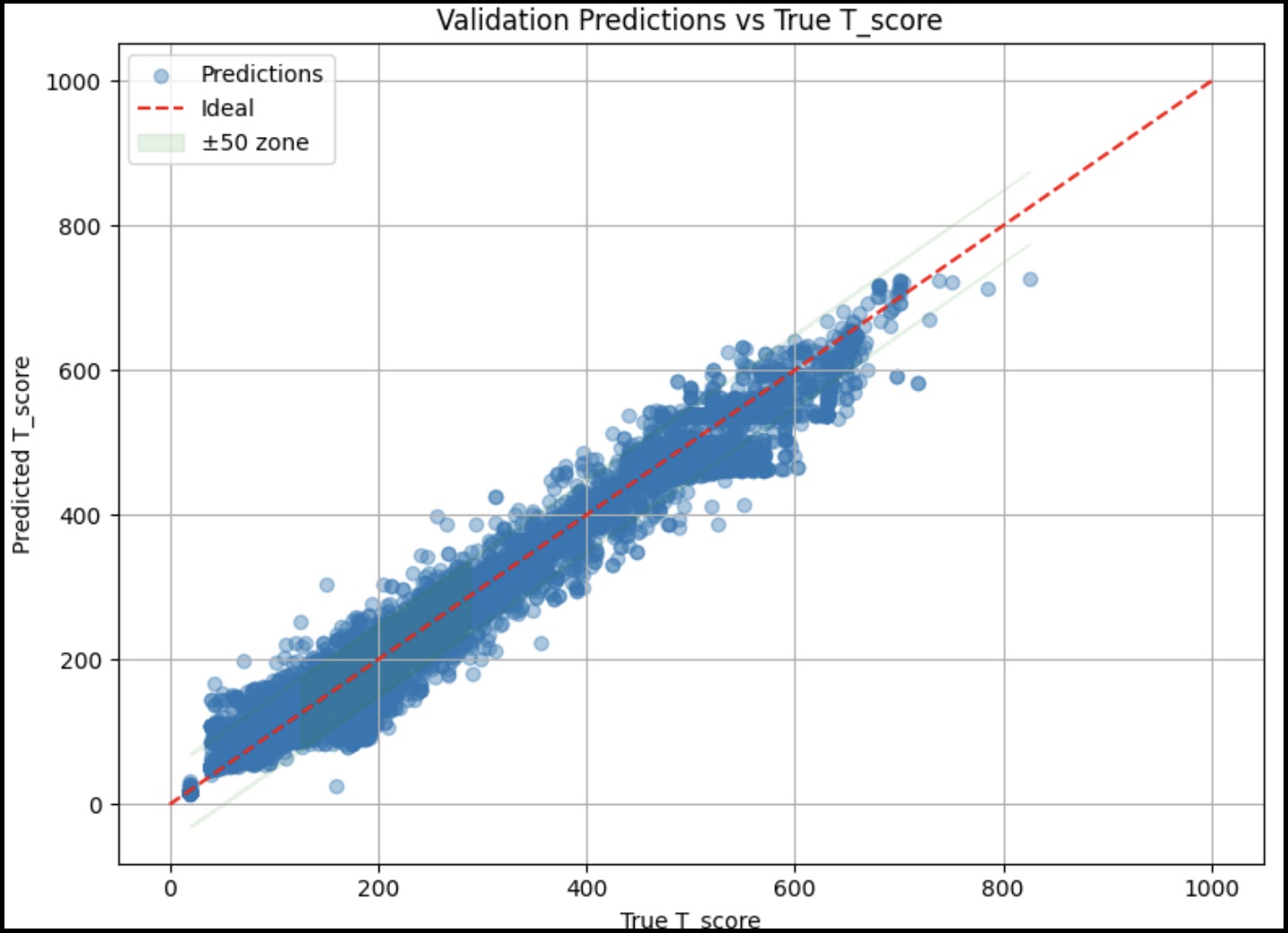}
    \end{minipage}
    \hfill
    \begin{minipage}[b]{0.48\textwidth}
        \centering
        \includegraphics[width=\textwidth]{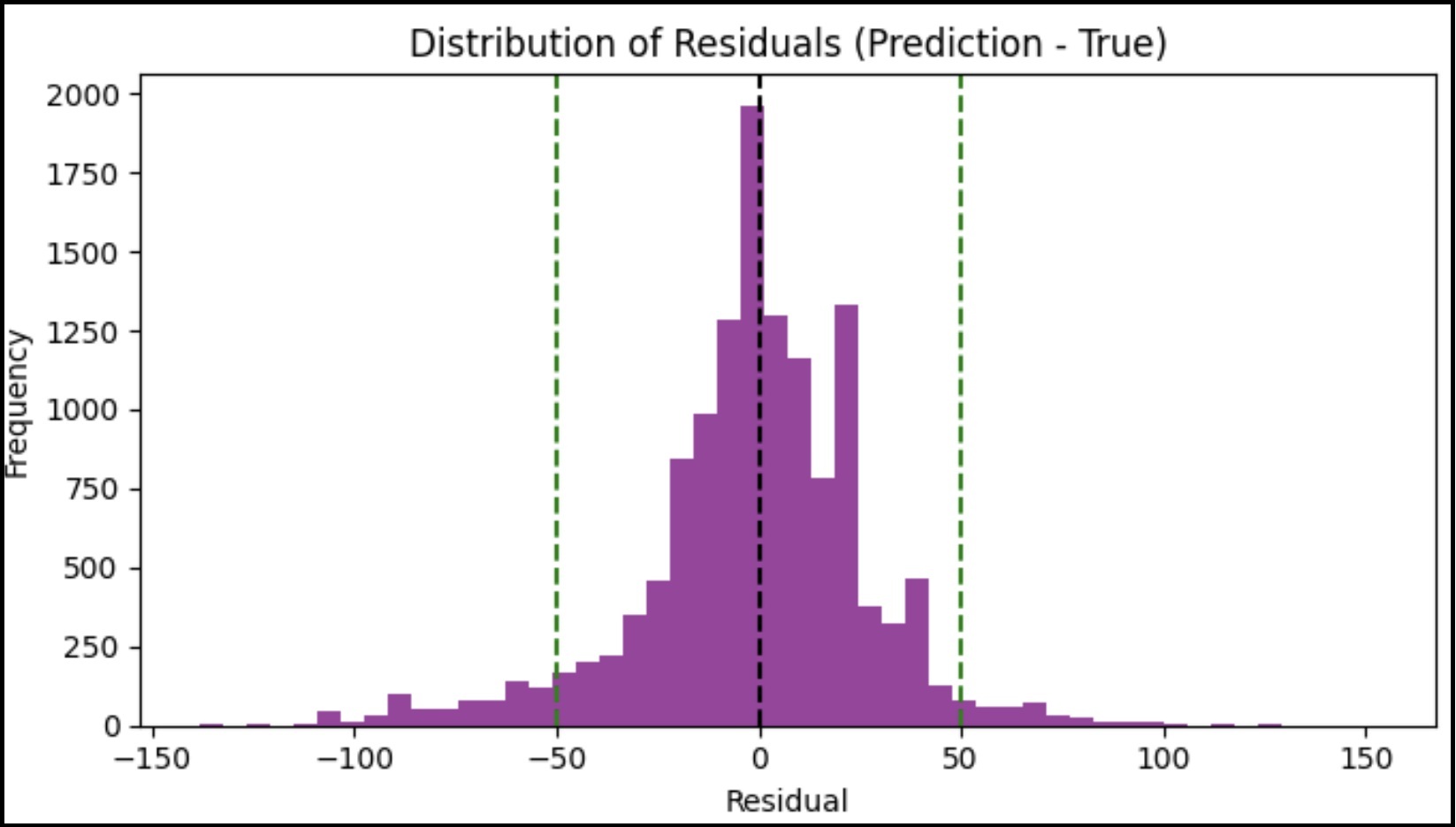}
    \end{minipage}
    \caption{(Left) Predicted vs. blueprint LP scores on the validation set. (Right) Distribution of residuals (prediction minus ground truth).}
    \label{fig:lp_score_results}
\end{figure}

To evaluate the effectiveness of our model in learning liquidity provider behavior, we compare the predicted T\_scores against blueprint-defined targets. Figure~\ref{fig:lp_score_results} presents two key visualizations: the left plot (a) shows the predicted versus ground truth T\_scores across the validation set, while the right plot (b) illustrates the distribution of residuals—computed as the difference between predicted and true scores.

The results indicate strong alignment between the model’s predictions and the blueprint scores, validating its ability to generalize behavioral patterns across unseen wallets. Notably, \textbf{91.79\%} of predictions fall within a $\pm50$ range of the corresponding ground truth, a threshold chosen to account for natural behavioral variation near scoring bin edges.

This tolerance accommodates the inherent ambiguity in user behavior without overfitting to hard boundaries. The model effectively learns smooth transitions between behavioral categories, preserving the ordinal nature of blueprint bins while introducing flexibility through data-driven learning.

The residual distribution reinforces this interpretation: most prediction errors cluster near zero, with a light-tailed spread suggesting well-calibrated approximation rather than random noise. Together, these results support the robustness and fairness of our hybrid scoring pipeline.

\subsection{Swap Score}

\begin{figure}[H]
    \centering
    \begin{minipage}[b]{0.48\textwidth}
        \centering
        \includegraphics[width=\textwidth]{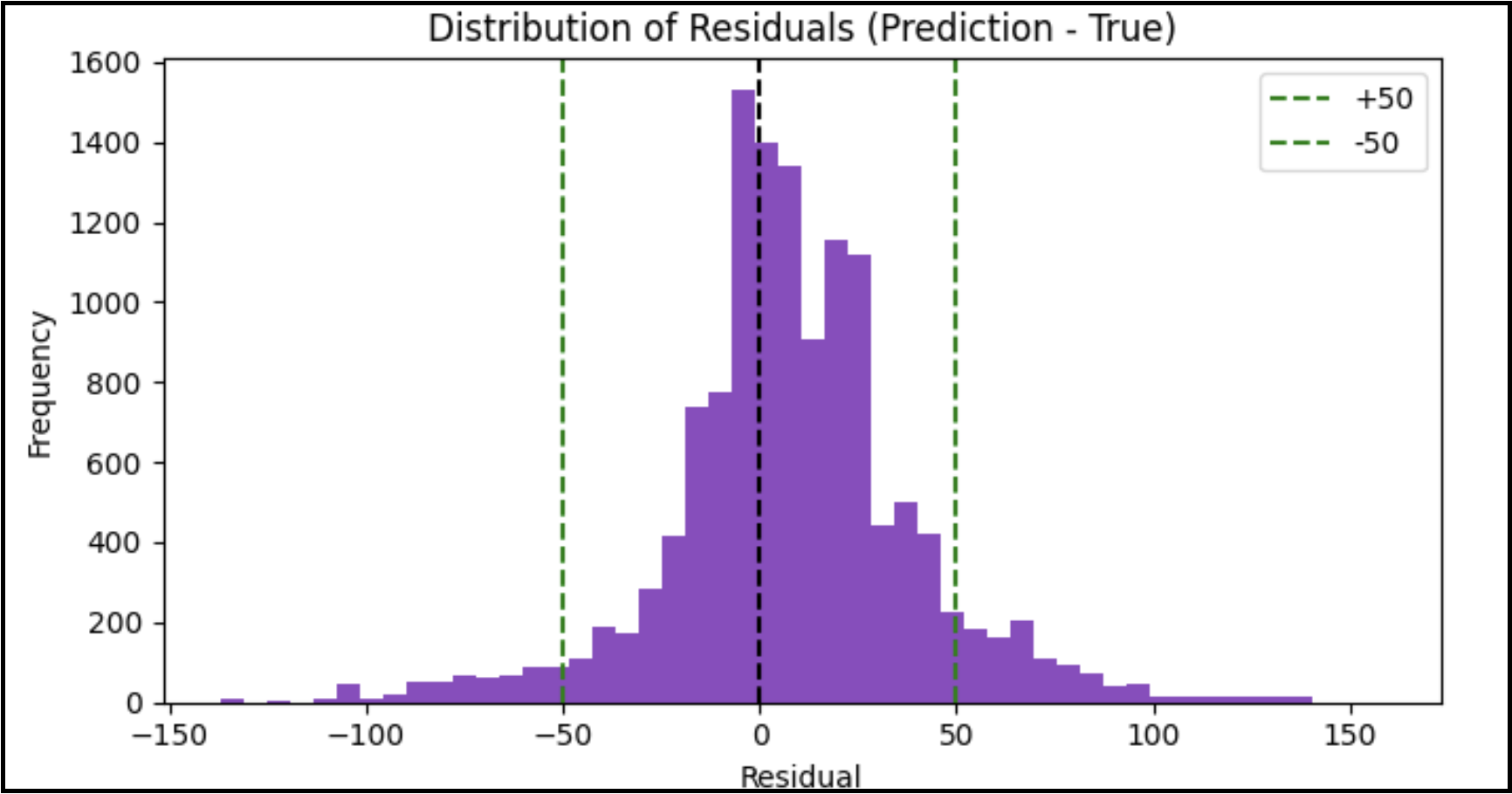}
    \end{minipage}
    \hfill
    \begin{minipage}[b]{0.48\textwidth}
        \centering
        \includegraphics[width=\textwidth]{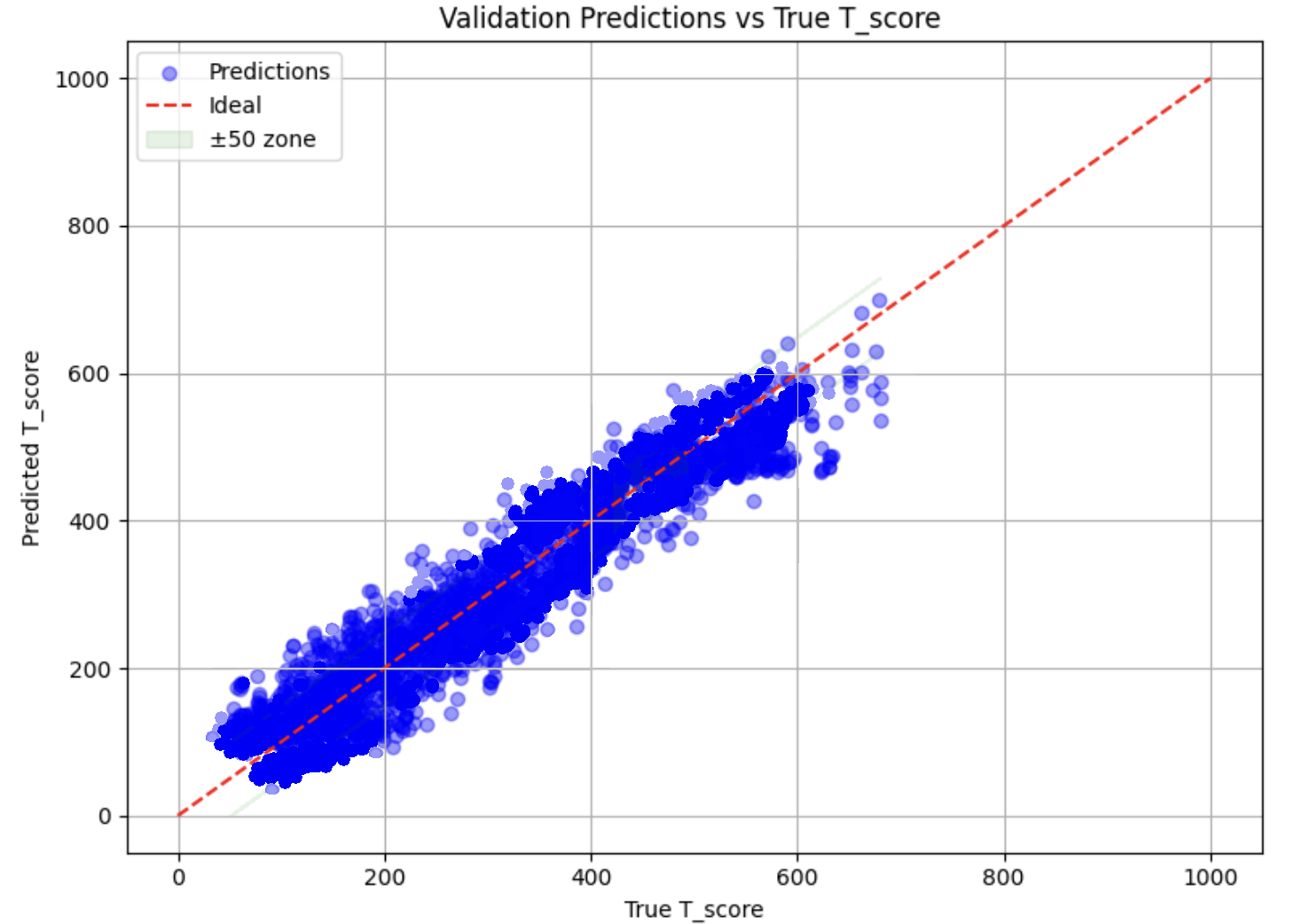}
    \end{minipage}
    \caption{(Left) Predicted vs. blueprint swap scores on the validation set. (Right) Distribution of residuals (prediction minus ground truth).}
    \label{fig:swap_score_results}
\end{figure}

We conduct a similar evaluation for swap behavior by comparing the model’s predicted T\_scores against blueprint-derived targets. Figure~\ref{fig:swap_score_results} shows the results: the left plot (a) illustrates predicted vs. ground truth scores, while the right plot (b) shows the distribution of residuals.

The model demonstrates strong performance, with \textbf{90.83\%} of predictions falling within a $\pm50$ window from the true score. This high agreement highlights the model’s ability to internalize diverse behavioral signals—such as swap frequency, pair volatility, holding time, and route integrity—while maintaining alignment with the blueprint’s ordinal intent.

The residual distribution is centered and compact, similar to that of LP scoring. This reflects a calibrated understanding of behavioral variability across swap types and confirms the model’s generalization to complex user profiles without overfitting.

\subsection{Feature-Score Correlation Analysis}
\subsubsection{Behavioral Trends Across LP Score Bins}

To further validate the alignment between LP scores and desirable liquidity provisioning behavior, we analyze average wallet behavior across score bins on a subset of the validation set. Table~\ref{tab:lp_bin_summary} summarizes key behavioral metrics across score ranges. For each bin, we report the average deposit and withdrawal volumes (in USD), the average holding duration (in days), the liquidity retention ratio (fraction of liquidity remaining in the pool), and the average monthly deposit frequency. These metrics provide a composite view of how behavior improves with increasing LP scores. 

Wallets in higher bins generally show lower withdrawal activity, higher liquidity retention, and significantly longer holding durations—indicating a stronger alignment with long-term, stable LP behavior. Additionally, we observe a bell-shaped distribution in wallet count across bins, with the majority of users concentrated in the middle score ranges (200–500). This further reinforces the discriminative nature of the scoring framework in capturing varying degrees of strategic participation. The final column denotes the number of wallets in each score bin. This analysis is based on a sample of \textbf{44,975} wallets.

\begin{table}[H]
\centering
\caption{Average LP behavior metrics across score bins (subset of validation set)}
\label{tab:lp_bin_summary}
\begin{tabularx}{\linewidth}{l *{6}{>{\raggedleft\arraybackslash}X}}
\toprule
\textbf{Score Range} & \textbf{Avg Deposit (\$)} & \textbf{Avg Withdraw (\$)} & \textbf{Avg Holding (days)} & \textbf{Liquidity Remaining} & \textbf{Deposit Frequency} & \textbf{Wallet Count} \\
\midrule
{[0, 100)}     & 185      & 1.35~K     & 1.87     & 0.00  & 27.68   & 4298  \\
{[100, 200)}   & 43.48~K  & 69.60~K    & 36.74    & 0.02  & 1.72~K  & 4799  \\
{[200, 300)}   & 153.53~K & 209.41~K   & 75.95    & 0.04  & 7.39~K  & 13781 \\
{[300, 400)}   & 1.55~M   & 1.66~M     & 157.72   & 0.08  & 3.22~K  & 12214 \\
{[400, 500)}   & 7.57~M   & 8.17~M     & 277.50   & 0.34  & 104.44  & 5451  \\
{[500, 600)}   & 840.75~K & 797.64~K   & 532.33   & 0.79  & 60.95   & 3284  \\
{[600, 700)}   & 557.73~K & 252.12~K   & 603.49   & 0.91  & 132.68  & 956   \\
{[700, 800)}   & 366.71~K & 822        & 575.48   & 0.98  & 0.46    & 171   \\
{[800, 900)}   & 99.52~K  & 422        & 347.48   & 0.99  & 0.48    & 16    \\
{[900, 1000)}  & 77.89~K  & 0.93       & 909.02   & 1.00  & 0.12    & 5     \\
\bottomrule
\end{tabularx}
\end{table}

\subsubsection{Behavioral Trends Across Swap Score Bins}

To evaluate how well the swap score reflects trading discipline and engagement, we analyze aggregated wallet behavior across score bins on a subset of the validation set. Table~\ref{tab:swap_bin_summary} reports the average swap volume (in USD), holding time between swaps (in days), total swap count, average number of unique tokens interacted with, and wallet count per bin. These metrics offer insight into how behavioral intensity varies across the scoring spectrum.

Higher scoring wallets consistently demonstrate stronger trading engagement, with significantly higher swap volumes, increased transaction frequency, and greater diversity in token interactions. In contrast, lower score bins are characterized by sparse activity—few swaps, low volumes, and limited pool exposure—indicating exploratory or inconsistent participation.

While holding duration is not strictly monotonic, the highest scoring wallets dominate in both total volume and frequency. Notably, the number of unique tokens traded increases with score, suggesting that top-tier users interact across multiple pools and trading pairs. This reflects strategic depth, pool coverage, and conviction in participation.

Overall, the score successfully differentiates between low-intent users and sophisticated swappers, reinforcing its utility in classifying on-chain trading behavior. This analysis is based on a sample of \textbf{33,491} wallets.

\begin{table}[H]
\centering
\footnotesize
\caption{Average behavioral metrics across swap score bins}
\label{tab:swap_bin_summary}
\begin{tabular}{lrrrrr}
\toprule
\textbf{Score Bin} & \textbf{Avg Volume (\$)} & \textbf{Avg Holding (days)} & \textbf{Avg Swap Count} & \textbf{Avg Unique Tokens} & \textbf{Wallet Count} \\
\midrule
{[0, 100)}   & 4.06~K     & 1.00   & 1.07   & 2.00   & 5554  \\
{[100, 200)} & 36.21~K    & 3.87   & 1.76   & 2.26   & 12859 \\
{[200, 300)} & 95.30~K    & 28.79  & 4.18   & 2.63   & 8275  \\
{[300, 400)} & 277.80~K   & 43.47  & 9.78   & 2.69   & 4633  \\
{[400, 500)} & 1.17~M     & 23.78  & 24.43  & 2.91   & 1895  \\
{[500, 600)} & 8.63~M     & 11.90  & 158.63 & 3.19   & 959   \\
{[600, 700)} & 32.42~M    & 12.47  & 473.40 & 3.92   & 288   \\
{[700, 800)} & 82.44~M    & 8.48   & 1000.44& 6.88   & 18    \\
\bottomrule
\end{tabular}
\end{table}

Together, these results confirm the robustness and behavioral sensitivity of our scoring framework. The LP and swap score evaluations demonstrate clear alignment between higher scores and disciplined, value-aligned user behavior—such as long-term liquidity provision, low churn, high swap volume, and broad token coverage. Residual analysis validates the model’s ability to generalize blueprint logic across unseen wallets with minimal error, while bin-wise feature summaries provide intuitive proof that score increases reflect increasingly strategic participation.

Importantly, our analysis explicitly excludes dusk LPs and dusk swappers—wallets with only one observed transaction (e.g., a single deposit or swap) and typically negligible volume (often below $10 for LPs or $50 for swaps). These users lack meaningful behavioral history, rendering their inclusion uninformative for pattern learning and biasing score distribution toward inactivity. By filtering them out, we ensure that our results reflect behaviorally significant activity and the model's ability to differentiate meaningful user engagement across DeFi ecosystems.

These findings highlight the effectiveness of combining rule-based interpretability with data-driven refinement to derive reliable, context-aware wallet scores across decentralized exchange activity.
\subsection*{External Validation Report}

To supplement the scoring framework proposed in this work, we conducted a detailed case study across more than 50{,}000 Uniswap v3 wallets. The accompanying validation report, published at \url{https://github.com/Dhanashekar-k/dex}, provides in-depth behavioral analysis across score bins, category-wise LP and swap participation, and real-world wallet archetypes. It includes breakdowns of liquidity retention, volatility, holding time, fee tier usage, and swap routing patterns. This document serves as both a sanity check and a practical guide to interpreting the zScore outputs in the context of DeFi wallet behavior.

\section{Future Work}

While this paper introduces a robust and interpretable scoring framework tailored for Uniswap v3, the long-term vision extends far beyond protocol-specific scoring. A key direction for future work is the development of a unified behavioral reputation score that operates across multiple DeFi protocols and ecosystems.

Such a cross-protocol scoring system would allow us to quantify user reliability, intent, and strategic alignment holistically—whether they are providing liquidity on Uniswap, borrowing on Aave, repaying on Morpho, or swapping across DEX aggregators. Achieving this requires standardized behavioral feature schemas, normalization techniques that account for protocol-specific risk and scale, and multi-modal training on cross-protocol datasets.

A unified zScore would enable composable reputation across DeFi, making it possible to design protocol-agnostic incentive structures, shared risk models, and federated identity layers. This vision aligns with the broader movement toward on-chain trust systems that preserve privacy while unlocking cooperation in decentralized environments.

\section{Conclusion}

In this work, we present a robust and interpretable framework for behavioral scoring of DEX users, focusing on liquidity provision and token swap activity. Our proposed zScore emerges as a powerful unified metric that reflects both the quality and consistency of on-chain behavior, capturing nuances that simple volume- or frequency-based heuristics fail to represent.

By combining a domain-informed blueprint with an AI-driven refinement model, we achieve a balance between structure and flexibility. The blueprint enforces category-level contribution caps, ensuring that no single behavioral trait can dominate the score—a design choice that promotes well-rounded participation. The neural model, trained on noise-augmented scores, learns smooth transitions between behavioral clusters, adapting to pool-level context and real-world behavioral variance.

Empirical evaluations confirm that zScores align closely with strategic behavior patterns. High-scoring users exhibit strong liquidity retention, disciplined trading activity, and broad engagement across pools—indicating that the score meaningfully differentiates long-term, protocol-aligned participants from short-term or exploratory actors. Feature-to-score analysis further validates the score’s interpretability and effectiveness in ranking users across the behavioral spectrum.

Overall, zScore provides a scalable, interpretable, and adaptive measure of wallet quality for decentralized finance. It can serve as a foundational layer for downstream applications such as credit risk modeling, airdrop eligibility, governance weighting, and capital allocation in on-chain financial systems.



\begin{thebibliography}{10}

\bibitem{malamud2017decentralized}
Semyon Malamud and Marzena Rostek.
\newblock Decentralized exchange.
\newblock {\em American Economic Review}, 107(11):3320--3362, 2017.

\bibitem{lo2021uniswap}
Yuen~C Lo and Francesca Medda.
\newblock Uniswap and the emergence of the decentralized exchange.
\newblock {\em Journal of financial market infrastructures}, 10(2):1--25, 2021.

\bibitem{schueffel2019evaluating}
Patrick Schueffel and Nikolaj Groeneweg.
\newblock Evaluating crypto exchanges in the absence of governmental frameworks-a multiple criteria scoring model.
\newblock {\em Available at SSRN 3432798}, 2019.

\bibitem{fantazzini2021crypto}
Dean Fantazzini and Raffaella Calabrese.
\newblock Crypto exchanges and credit risk: Modeling and forecasting the probability of closure.
\newblock {\em Journal of Risk and Financial Management}, 14(11):516, 2021.

\bibitem{aigner2021uniswap}
Andreas~A Aigner and Gurvinder Dhaliwal.
\newblock Uniswap: Impermanent loss and risk profile of a liquidity provider.
\newblock {\em arXiv preprint arXiv:2106.14404}, 2021.

\bibitem{lin2024riskprop}
Dan Lin, Jiajing Wu, Qishuang Fu, Zibin Zheng, and Ting Chen.
\newblock Riskprop: Account risk rating on ethereum via de-anonymous score and network propagation.
\newblock {\em IEEE Transactions on Dependable and Secure Computing}, 2024.

\bibitem{nguyen2024reputation}
Mau-Tra Nguyen, Tuan-Dat Trinh, and Viet-Bang Pham.
\newblock A reputation scoring framework for lending protocols using the pagerank algorithm.
\newblock In {\em International Symposium on Information and Communication Technology}, pages 478--494. Springer, 2024.

\bibitem{packin2024decentralized}
Nizan~Geslevich Packin and Yafit Lev-Aretz.
\newblock Decentralized credit scoring: Black box 3.0.
\newblock {\em American Business Law Journal}, 61(2):91--111, 2024.

\bibitem{shimpi2024credit}
Mayuri Shimpi.
\newblock {\em Credit Score-Based Lending System On the Ethereum Platform}.
\newblock PhD thesis, San Jos{\'e} State University, 2024.

\bibitem{mussoi2025risk}
Lucas Mussoi~Almeida, Fernanda~Maria M{\"u}ller, and Marcelo~Scherer Perlin.
\newblock Risk forecasting comparisons in decentralized finance: An approach in constant product market makers.
\newblock {\em Computational Economics}, 65(1):395--428, 2025.

\bibitem{heimbach2022risks}
Lioba Heimbach, Eric Schertenleib, and Roger Wattenhofer.
\newblock Risks and returns of uniswap v3 liquidity providers.
\newblock In {\em Proceedings of the 4th ACM Conference on Advances in Financial Technologies}, pages 89--101, 2022.

\bibitem{carter2021defi}
Nic Carter and Linda Jeng.
\newblock Defi protocol risks: The paradox of defi.
\newblock {\em Regtech, suptech and beyond: innovation and technology in financial services” riskbooks--forthcoming Q}, 3, 2021.

\bibitem{doerr2021defi}
Jon~Frost Doerr, Anneke Kosse, Asad Khan, Ulf Lewrick, Beno{\^\i}t Mojon, Benedicte Nolens, and Tara Rice.
\newblock Defi risks and the decentralisation illusion.
\newblock {\em BIS Quarterly Review}, 21, 2021.

\bibitem{bucker2022transparency}
Michael B{\"u}cker, Gero Szepannek, Alicja Gosiewska, and Przemyslaw Biecek.
\newblock Transparency, auditability, and explainability of machine learning models in credit scoring.
\newblock {\em Journal of the Operational Research Society}, 73(1):70--90, 2022.

\bibitem{provenzano2020machine}
Angela~Rita Provenzano, Daniele Trifiro, Alessio Datteo, Lorenzo Giada, Nicola Jean, Andrea Riciputi, G~Le Pera, Maurizio Spadaccino, Luca Massaron, and Claudio Nordio.
\newblock Machine learning approach for credit scoring.
\newblock {\em arXiv preprint arXiv:2008.01687}, 2020.

\bibitem{udupi2025zscore}
Himanshu Udupi, Ashutosh Sahoo, Parag Paul, Petrus~C Martens, et~al.
\newblock zscore: A universal decentralised reputation system for the blockchain economy.
\newblock {\em arXiv preprint arXiv:2503.05718}, 2025.

\bibitem{jain2024wire}
Suraj~Shamsundar Jain, Huancheng Zhou, and Guofei Gu.
\newblock Wire: Web3 integrated reputation engine.
\newblock In {\em 2024 IEEE 44th International Conference on Distributed Computing Systems (ICDCS)}, pages 1388--1399. IEEE, 2024.

\bibitem{aspris2021decentralized}
Angelo Aspris, Sean Foley, Jiri Svec, and Leqi Wang.
\newblock Decentralized exchanges: The “wild west” of cryptocurrency trading.
\newblock {\em International Review of Financial Analysis}, 77:101845, 2021.

\bibitem{hagele2024centralized}
Sascha H{\"a}gele.
\newblock Centralized exchanges vs. decentralized exchanges in cryptocurrency markets: A systematic literature review.
\newblock {\em Electronic Markets}, 34(1):33, 2024.

\bibitem{dastile2020statistical}
Xolani Dastile, Turgay Celik, and Moshe Potsane.
\newblock Statistical and machine learning models in credit scoring: A systematic literature survey.
\newblock {\em Applied Soft Computing}, 91:106263, 2020.

\bibitem{dumitrescu2022machine}
Elena Dumitrescu, Sullivan Hu{\'e}, Christophe Hurlin, and Sessi Tokpavi.
\newblock Machine learning for credit scoring: Improving logistic regression with non-linear decision-tree effects.
\newblock {\em European Journal of Operational Research}, 297(3):1178--1192, 2022.

\end{thebibliography}
\end{document}